\documentclass{ws-procs9x6}

\begin{document}

\title{Charge Fluctuation in Heavy Ion Collisions
\footnote{\uppercase{T}alk given at the \uppercase{S}ymposium on
\uppercase{M}ultiparticle
\uppercase{D}ynamics, 
\uppercase{A}lushta, 
\uppercase{U}kraine, 
\uppercase{S}ept. 2002}
}

\author{\underbar{Fritz W. Bopp\footnote{\uppercase{P}artial support from the INTAS
grant 97-31696 is acknowledged.} } and Johannes Ranft}

\address{Universit{\"a}t Siegen, Fachbereich Physik, D--57068 Siegen, Germany }


\maketitle

\abstracts{
Charge fluctuations observed in early fixed-target proton-proton experiments
are consistent with string models. In central heavy ion events the picture can
change in two ways: strings can interact and find new ways to hadronize
or they can be effectively inactivated to lose their dynamical role as ordering
mechanism. Widely different charge fluctuations can be expected. The dispersion
of the charges in a central rapidity box is an advantageous measure. In an explicit
Dual-Parton-Model calculation using the DPMJET code and a randomized modification
to simulated charge equilibrium, various energies and different nuclear sizes
were considered. Local fluctuations were found to be a serious problem.
However,
for large enough detection regions charged particle fluctuations can provide
a clear signal reflecting the basic dynamics of central heavy ion processes. 
}

\subsection*{{\small Charge Fluctuations in Fixed Target Hadron-Hadron Experiments }\small }

{\small At fixed target experiments in hadronic multi-particle production it
was possible to measure all charges of forward particles. In this way significant
results could be obtained even with low energies available at the seventies\cite{idschok73}:}
\vspace{-0.6em}\begin{itemize}
\item{\small The charge fluctuations involve a restricted rapidity range. }
\item{\small Qualitative agreement was  obtained the Quigg-Thomas relation. }{\small \par}
\end{itemize}
\vspace{-0.6em}{ {\small The Quigg-Thomas relation\cite{quigg73} was initially
based on intermediately produced neutral clusters \cite{bopp78,ranft75}. It
postulates for charge fluctuation across a rapidity \( y \) boundary 
\begin{equation}
\label{eq-1}
<\delta Q_{>y}^{2}>=<(Q_{>y}-<Q_{>y}>)^{2}>=c\cdot dN^{\mathrm{non}\, \mathrm{leading}}_{\mathrm{charge}}/dy
 . \end{equation}
To quantitatively fit the constant \( c \) with known resonances links with
\( q \) resp. \( \bar{q} \) exchanges had to be added\cite{baier74}. }\small \par}

{\small Such links appear in string models. We re-checked this old result using
the Dual Parton Model code DPMJET\cite{DPMJET}: For pp-scattering at laboratory energies
of 205 GeV good agreement is obtained. }

\subsection*{{\small Strings as Ordering Mechanism }\small }

{\small In QED real and virtual soft or collinear emissions cancel as the final
states cannot be distinguished in measurement. For QCD such contributions involve
long time scales leaving the perturbative regime. With the comparatively compact
hadronic final states the emissions now cancel as the final states are equal. }{\small \par}

{\small In string models the hadronic final state is thought to be composed
of color singlets called strings. 
If different
soft or collinear contributions to a string production amplitude are summed
their phases can lead to cancellations. 
In this way strings can act as infrared regulators. }{\small \par}

{\small It is possible that the usual soft phenomenology would emerge as extension
of PQCD, if these cutoffs could be properly implemented. 
Strings can play an essential role as ordering mechanism of the dynamics 
of their production.}{\small \par}

\subsection*{{\small What changes for Heavy Ion Scattering?}\small }

{\small With more interactions per nucleon strings will get more numerous and
shorter. There are two possible quite distinct consequences:}{\small \par}
\vspace{-0.6em}
\begin{itemize}
\item {\small Denser} \textbf{\emph{\small strings should interact}} {\small and find
a different, possibly more efficient way to hadronize.}{\small \par}
\item {\small A very large number of interactions can be expected to  essentially} \textbf{\emph{\small destroy
the strings}} {\small as infrared regulator or ordering mechanism.
The ensemble needed to describe the scattering  then involves a much larger number of states. }{\small \par}
\end{itemize}
\vspace{-0.6em}
{\small Reasonable expectations for both cases are respectively:}{\small \par}

{\small a} \textbf{\emph{\small reduction}} {\small in density, an increase
in baryon pairs and  in strangeness\cite{amelin01}, }{\small \par}

{\small an} \textbf{\emph{\small increase}} {\small in density, possibly looking
like local thermalization.}{\small \par}

{\noindent {\small RHIC data seem to favor the first option. Unfortunately
explicit models show large uncertainties. Clarification can
come from charge fluctuation measurements.}\small \par}

\subsection*{{\small Charge Fluctuations in Heavy-Ion Scattering Experiments}\small }

{\small In heavy ion experiments the charge distribution of the particle contained
in a central box with a given rapidity range \( [-y_{\mathrm{max}.},+y_{\mathrm{max}.}] \)
can be measured and the dispersion of this distribution \( <\textrm{ }\delta Q^{2}> \)
can be obtained to sufficient accuracy. In comparison to the fluctuations in
the forward backward charge distributions the charge distribution into a central
box (having to have two sufficiently separated borders) can be expected to require
roughly twice the rapidity range to obtain information about long range charge
flow.}{\small \par}

{\small Within the framework of equilibrium models the dispersion  was proposed 
to distinguish between particles emerging from an equilibrium quark-gluon
gas or from an equilibrium hadron gas\cite{heinz00,bleicher}. It should be
pointed out that this estimate is not without theoretical problems \cite{bopp81,bopp01}
having to do with the hadronisation process. }{\small \par}

\vspace{0.3cm}
{\small Besides the classic charge dispersion 
\begin{equation}
\label{eq-2}
<\delta Q^{2}>=<(Q-<Q>)^{2}>
\end{equation}
}{\small \par}

\noindent{{\small where \( Q=N_{+}-N_{-} \) is the net charge inside
the box, it was proposed to just measure the mean stan\-dard deviation of the
ratio \( R \) of positive to negative particles or the ratio \( F \) of the
net charge to the total number of charged particles in the box. The motivation
for choosing these ratios was to reduce the dependence of multiplicity fluctuations
caused by the event structure. These quan\-ti\-ties have problems for
hadron-hadron
or non-central heavy-ion events\cite{bopp01,phenix02}. As any conclusion will have to
de\-pend on a comparison of central processes with minimum bias and proton-proton
events, there is a clear advantage to stick to the dispersion of the net charge
distribution.}\small \par}

{\small The \( \phi  \) - and \( \Gamma  \) - measures also considered\cite{voloshin} are closely related to \( <\delta Q^{2}> \) .}{\small \par}

\subsection*{{\small Quark Line Structure and Fluctuations
in the Charge Flow}\small }

{\small To visualize the meaning of charge flow measurements it is helpful to
introduce the hypothesis that the flavor distribution of individual quarks
factorizes. It is an adequate approximation, especially if long range fluctuations
are considered. }{\small \par}

{\small The hypothesis leads to a} \emph{\small generalization
of the Quigg-Thomas relation}{\small \cite{baier74} determining  the 
correlation of the charges exchanged  across two  arbitrary boundaries.
A combination of such correlations yields the fluctuation of the charge 
within a \( [-y_{\mathrm{max}.},+y_{\mathrm{max}.}] \)
box:
\begin{equation}
\label{eq-3}
<\delta Q[\mathrm{box}]^{2}>=n^{\mathrm{quark}\:\mathrm{lines}}_{\mathrm{entering}\:
\mathrm{box}}<(\delta q)^{2}>
\end{equation}
}\noindent \small {where \(  Q[\mathrm{box}] \) is the  charge in the box,
where \( n^{\mathrm{quark}\:\mathrm{lines}}_{\mathrm{entering}\: \mathrm{box}} \) is the number
of quark lines entering or leaving the box, and 
where  \( q \) is the charge of the quark on such a line.
With the  notation: \(  \delta Q = Q - <Q>\),  
values  \( <\delta q^{2}> = 0.22\cdots 0.25 \)
can be  obtained.
 }{\small \par}

{\small The relation allows to easily evaluate simple situations like the} \emph{\small thermalized
limit} {\small of a small box with an infinite reservoir outside. In an ``}\emph{\small hadron
gas case}{\small '' all particles contain two independent quarks coming from outside;
in the so-called ``}\emph{\small quark gluon gas case}{\small '' one of the quarks
of each meson comes from the outside the other is ignored as a local hadronisation
affair (one ignores \( <q>\ne 0 \) ). }{\small \par}

\subsection*{{\small Expanding Box}\small }

{\small \emph {For a tiny box} ---  considering only at the first order in \( \Delta y \) ---
one trivially obtains the hadron gas value \( <\delta Q^{2}>/<N_{\mathrm{charged}}>=1
\).} \small{ 
If the box size increases to} \emph{\small one or two
units of rapidity} {\small on each side this ratio will decrease, as realistic
models contain a short range component in the charge fluctuations. }{\small \par}

{\small After a box size passed the short range the} \emph{\small decisive region}
{\small starts. In all global equilibrium models the ratio will have to reach
a flat value.  }{\small \par}

{\small If a} \emph{\small large box} {\small involves a significant part of
the phase space the overall charge conservation has to be considered with a 
correction factor \( \propto 1-y_{\mathrm{max}.}/Y_{\mathrm{kin}.\mathrm{max}.} \).
At present energies the decisive and large region are not separated.}{\small \par}

\subsection*{{\small String Model Predictions}\small }

{\noindent \emph{\small Charges are locally compensated} {\small as the
range spanned by quark lines in links or during resonance decays is limited.
The total contribution will be determined by the density of quark lines reflecting
the} \emph{\small number of strings} {\small at the boundaries:
\begin{equation}
\label{eq-4}
<\delta Q^{2}>\, \propto \rho _{\mathrm{charged}}(y_{\mathrm{max}.}).
\end{equation}
 }\small \par}

{\small This resulting scaling is illustrated in a DPMJET\cite{DPMJET} comparison between both quantities
in (4) shown in Fig.\- 1  for RHIC and LHC energies.}
\begin{figure}
{\par\centering \resizebox*{!}{0.3\textheight}{\includegraphics{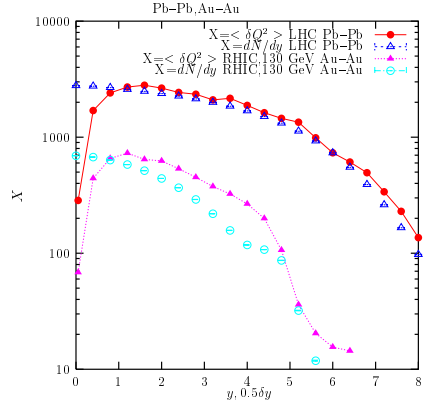}} \par}

\caption{{ The dispersion of the charge distribution and the density
on the box boundary for central gold gold resp. lead lead
scattering at RHIC and LHC energies.}\small }
\end{figure}
{\small For smaller boxes there is a correction as some of the quark lines 
intersect
both boundaries. For large rapidity sizes there is a minor increase from the
leading charge flow \( Q_{L} \) originating in the incoming 
particles\cite{Capella:1989gc}. In a
more careful consideration \cite{bopp78} one can
subtract this contribution \( <Q_{L}>(1-<Q_{L}>) \) and concentrate truly on
the fluctuation. A simple estimate --- with a width of neighboring string break
ups and a width from resonance decays --- leads to consistent values \cite{bopp01}. }{\small \par}

{\small A comparable result was obtained for the proton-proton case
\cite{bopp01}. }{\small \par}

\subsection*{{\small String Model versus  ``Hadron Gas'' }\small }

{\small It was argued \cite{bleicher} that the experimental results should
be ``purified'' to account for charge conservation. We prefer
a reference model with a posteriori} \emph{\small randomized charges}{\small .
This unbiased method can be obviously also directly applied to experimental
data.  Using DTMJET 
for RHIC and} \textbf{\small }{\small LHC energies for} \emph{\small proton-proton}
{\small and} \emph{\small central lead-lead} {\small collisions we obtain the
``statistical'' prediction shown in Fig.\- 2. }{\small \par}

{\small 
We employed the correction factor 
\( ( 1-\int _{0}^{y_{\mathrm{max}.}}\rho _{\mathrm{charge}}^{}\: dy ) \- 
/ \int _{0}^{Y_{\mathrm{kin}.\mathrm{max}.}}\rho _{\mathrm{charge}}^{}\: dy \)
proposed by \-\cite{bleicher} to check consistency and obtained expected the flat
distribution.}{\small \par}

\begin{figure}
{\par\centering \resizebox*{0.48\columnwidth}{0.27\textheight}{\includegraphics{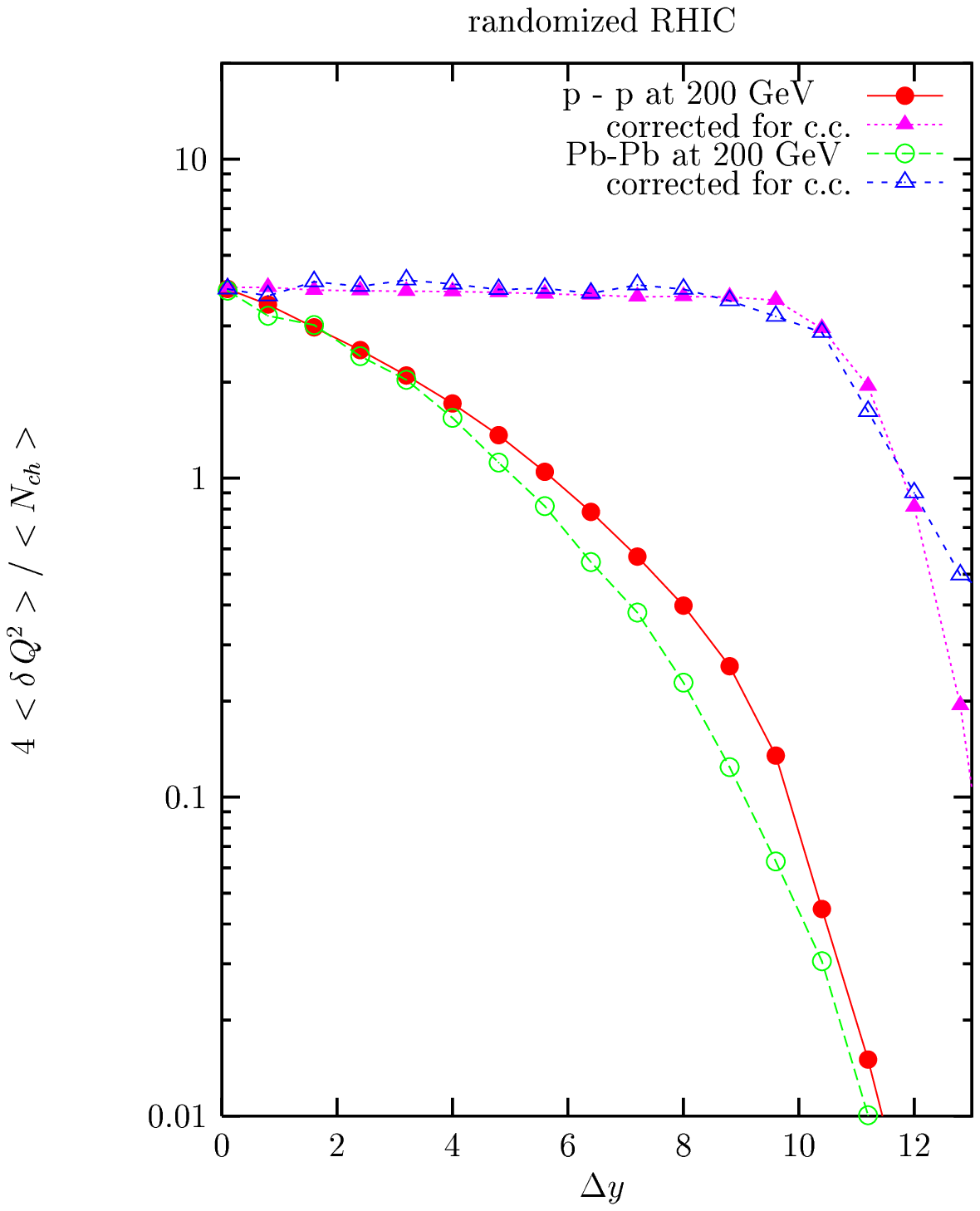}} 
\resizebox*{0.48\columnwidth}{0.27\textheight}{\includegraphics{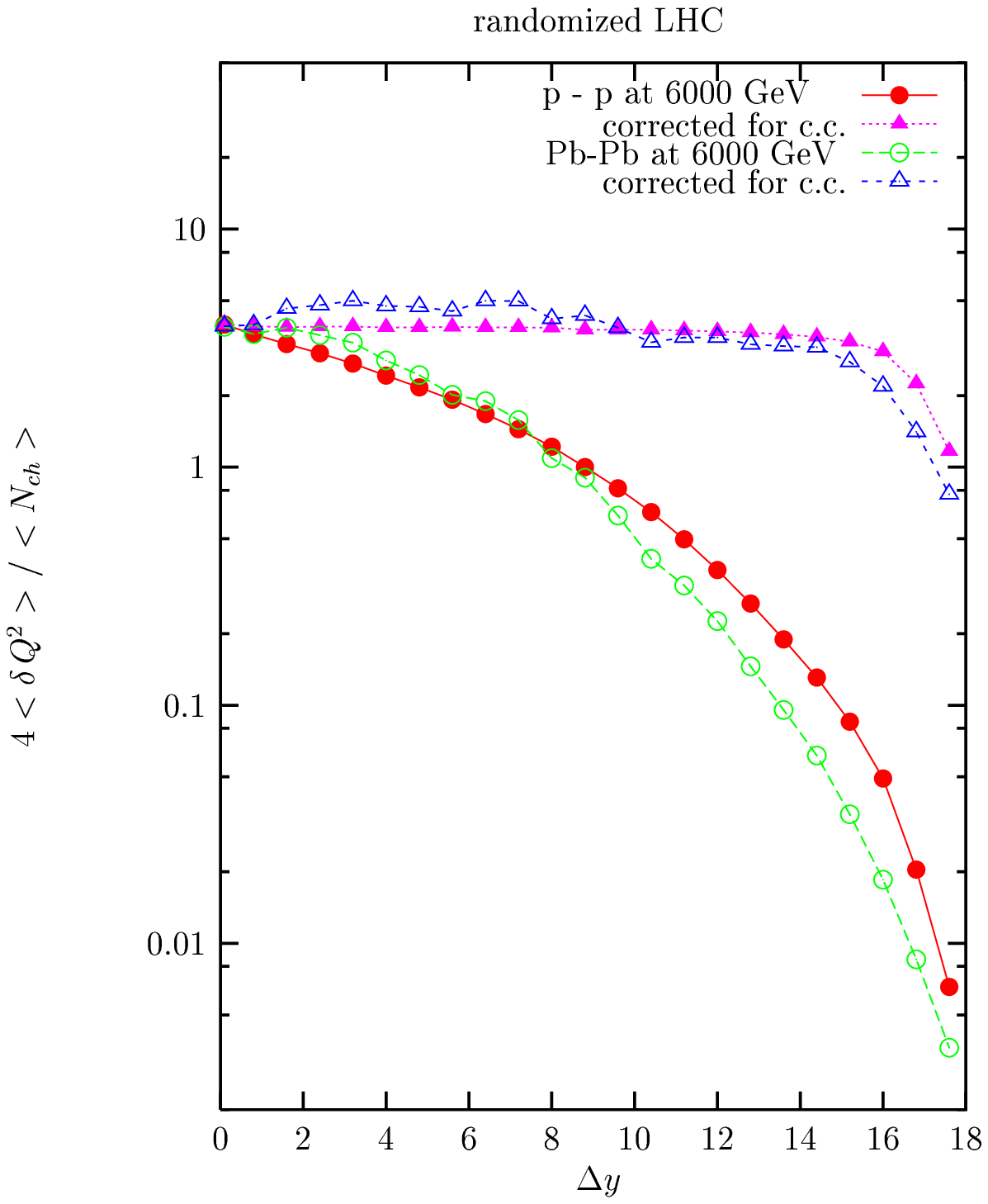}} \par}

\caption{{ Charge fluctuations with a posteriori randomized charges for p-p scattering
and the most central \protect\( 5\%\protect \) in Pb-Pb scattering at RHIC
energies (\protect\( \sqrt{s}=200\protect \) A GeV) and at LHC energies (\protect\( \sqrt{s}=6000\protect \)
A GeV). The results are also shown with a correction factor to account for the
overall charge conservation.}\small }
\end{figure}
\begin{figure}
{\centering \resizebox*{0.48\columnwidth}{0.25\textheight}{\includegraphics{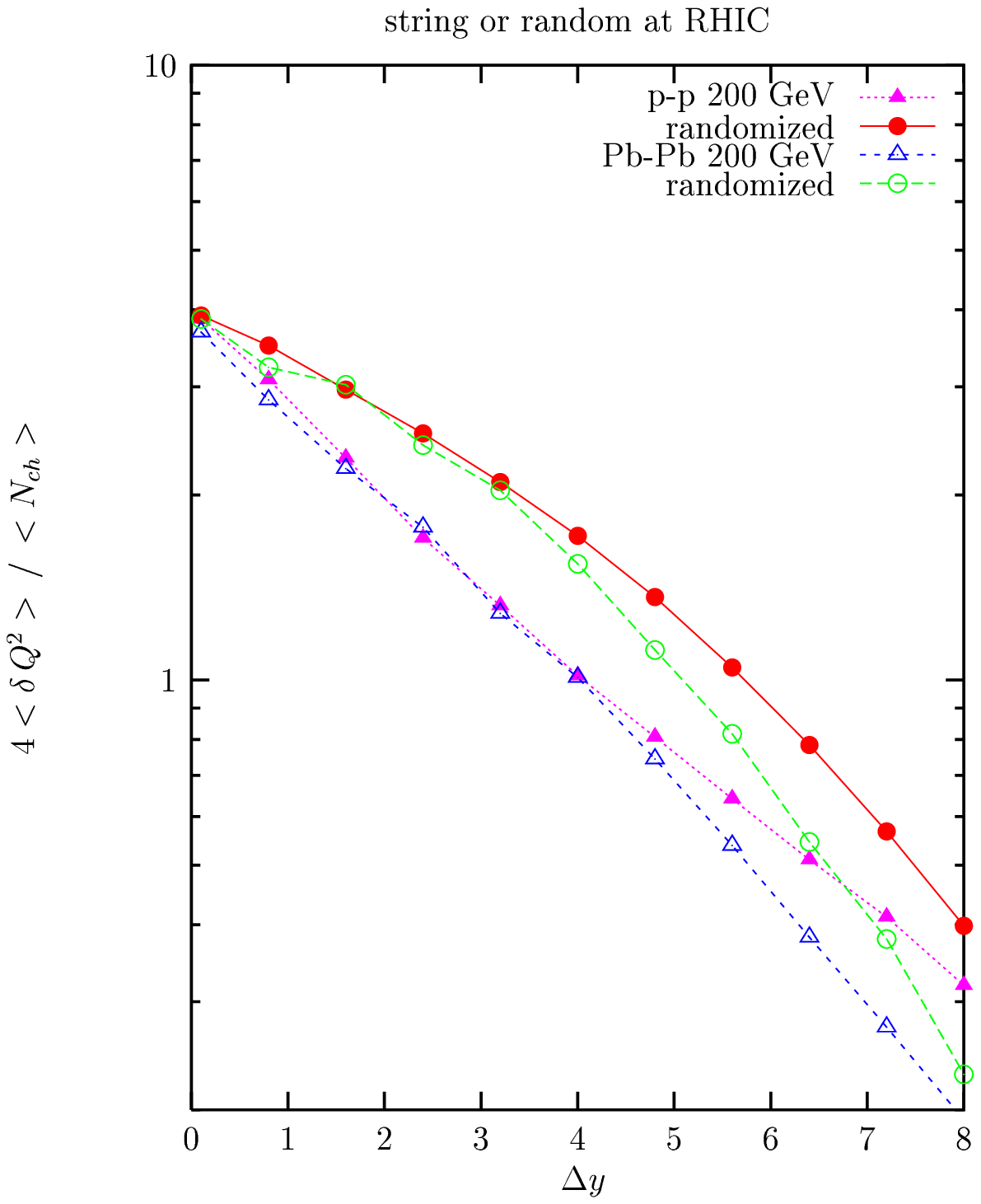}} 
\resizebox*{0.48\columnwidth}{0.25\textheight}{\includegraphics{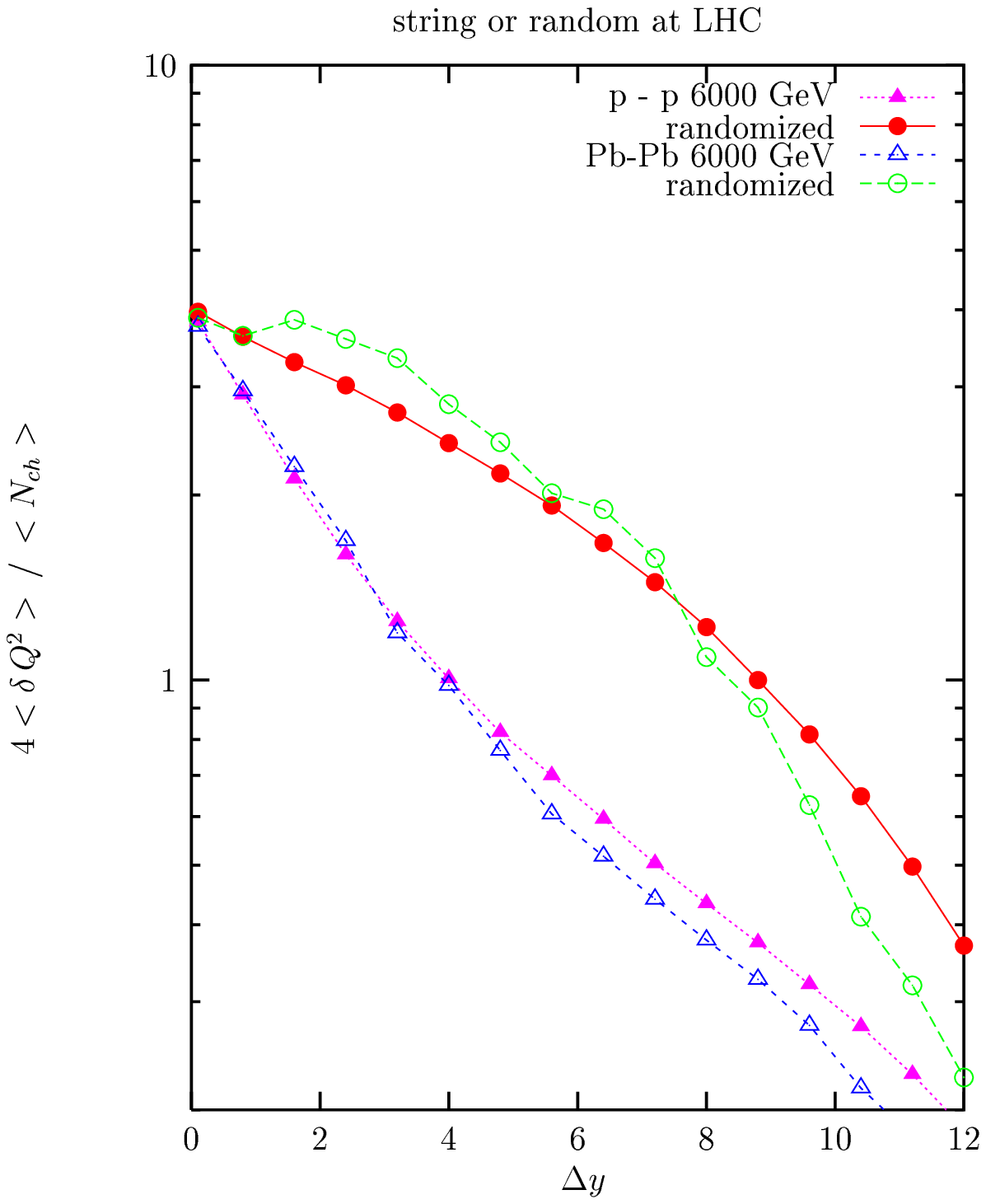}} \par}

\caption{{\small Comparison of the charge fluctuations obtained in a string
model DPMJET with a model using a posteriori randomized charges for
p-p scattering and the most central \protect\( 5\%\protect \) in
Pb-Pb scattering at RHIC energies (\protect\( \sqrt{s}=200\protect \)
A GeV) and at LHC energies (\protect\( \sqrt{s}=6000\protect \) A
GeV). }}
\end{figure}

{\small Taking the DPMJET string model and the randomized ``hadron gas'' version
as extreme cases the decisive power can be tested. As shown
in Fig.\-3 
there is a  \textbf{\em measurable} distinction at RHIC energies and
\textbf{\em sizable}
one at LHC energies.
 }{\small \par}

{\small The spectra change roughly by a
factor of 400 between simple proton-proton scattering and
central lead lead scattering. 
The suprising similarity between p-p and Pb-Pb in the Fig.\- 3 can 
be understood as collective effects to a large part not included in the model.
Also no dependence on the centrality was observed
in DPMJET for Pb-Pb scattering at RHIC energies (\( \sqrt{s}=200 \) A GeV)
\cite{bopp01} . 
This experimentally measurable centrality dependence allows 
to directly observe collective effects
without reference to a particular model.}{\small \par}

\section*{{\small Conclusion}\small }

{\small The dispersion of the charge distribution in a central box of varying
size is an extremely powerful measure. It allows to directly and quantitatively}
\emph{\small }{\small test} \textbf{\emph{\small the presence of equilibrizing
processes}} {\small and remaining} \textbf{\emph{\small dynamical corrections}}
{\small to equilibrized distributions.}{\small \par}


\begin{thebibliography}{10}
\bibitem{idschok73}{\small U.~Idschok et al. [Bonn-Hamburg-Munich Collaboration], Nucl.\ Phys.\ {\bf B67}, 93 (1973), A.~Bialas, K.~Fialkowski, M.~Jezabek and M.~Zielinski, Acta Phys.\ Polon.\ B {\bf 6}, 39 (1975),
J.~Whitmore, Phys.\ Rept.\ {\bf 27}, 187 (1976), L.~Foa, Phys.\ Rept.\ {\bf 22}, 1 (1975),
R.~Brandelik {\it et al.} [TASSO Collaboration], Phys.\ Lett.\ B {\bf 100}, 357 (1981), C.~Quigg, Phys.\ Rev.\ {\bf D12}, 834 (1975).}{\small \par}
\bibitem{quigg73}{\small C.~Quigg and G.~H.~Thomas, Phys.\ Rev.\ {\bf D7},2757 (1973), C.~Quigg, Phys.\ Rev.\ {\bf D12}, 834 (1975). }{\small \par}
\bibitem{bopp78}{\small F.~W.~Bopp, Riv.\ Nuovo Cim.\ {\bf 1}, 1 (1978). }{\small \par}
\bibitem{ranft75}{\small J.~Ranft, Fortsch.\ Phys.\ {\bf 23}, 467 (1975), S.~P.~Misra and B.~K.~Parida, Pramana{\bf 20}, 375 (1983),
S.~P.~Misra and B.~K.~Parida, Pramana{\bf 20}, 375 (1983), T.~T.~Chou and C.~N.~Yang, ~Phys.\ Rev.\ {\bf D7}, 1425 (1973),
J.~L.~Newmeyer and D.~Sivers, Phys.\ Rev.\ D {\bf 10}, 204 (1974), K.~F.~Loe, K.~K.~Phua and S.~C.~Chan, Lett.\ Nuovo Cim.\ {\bf 18}, 137 (1977),
E.~N.~Argyres and C.~S.~Lam, Phys.\ Rev.\ D {\bf 16}, 114 (1977), C.~B.~Chiu and K.~Wang, Phys.\ Rev.\ D {\bf 13}, 3045 (1976),
J.~Dias de Deus and S.~Jadach, Phys.\ Lett.\ B {\bf 66}, 81 (1977), M.~Jezabek, Phys.\ Lett.\ B {\bf 67}, 292 (1977). }{\small \par}
\bibitem{baier74}{\small R.~Baier and F.~W.~Bopp, Nucl.\ Phys.\ {\bf B79}, 344(1974), P.~Aurenche and F.~W.~Bopp, Nucl.\ Phys.\ {\bf B119}, 157 (1977).}{\small \par}
\bibitem{DPMJET}{\small J.~ Ranft, Phys.\ Rev. {\bf D 510} 64 (1995); J.~Ranft, hep-ph/9911213 (Siegen preprint SI-99-5); J.~Ranft, hep-ph/9911232 (Siegen preprint SI-99-6).}{\small \par}
\bibitem{amelin01}{\small N.~S.~Amelin, N.~Armesto, C.~Pajares and D.~Sousa, Eur.\ Phys.\ J.\ C {\bf 22}, 149 (2001) [arXiv:hep-ph/0103060]. }{\small \par}
\bibitem{heinz00}{\small M.~Asakawa, U.~Heinz and B.~Muller, Phys.\ Rev.\ Lett.\ {\bf 85}, 2072 (2000),
S.~Jeon and V.~Koch, Phys.\ Rev.\ Lett.\ {\bf 85}, 2076 (2000) [hep-ph/0003168],
S.~Jeon and V.~Koch, Phys.\ Rev.\ Lett.\ {\bf 83}, 5435 (1999)[nucl-th/9906074].}{\small \par}
\bibitem{bleicher}{\small M.~Bleicher, S.~Jeon and V.~Koch, Phys.\ Rev.\ {\bf C62}, 061902 (2000) [hep-ph/0006201];
V.~Koch, M.~Bleicher and S.~Jeon, nucl-th/0103084, M.~Bleicher {\it et al.}, J.\ Phys.\ {\bf G25} (1999) 1859 [hep-ph/9909407]. }{\small \par}
\bibitem{bopp81}{\small F.~W.~Bopp,Nucl.\ Phys.\ {\bf B191}, 75 (1981).}{\small \par}
\bibitem{bopp01}{\small F.~W.~Bopp and J.~Ranft, 
Acta Phys.\ Polon.\ B {\bf 33}, 1505 (2002) [arXiv:hep-ph/0204010]. F.~W.~Bopp and J.~Ranft, Eur.\ Phys.\ J.\ C {\bf 22}, 171 (2001) arXiv:hep-ph/0105192, K.~Fialkowski and R.~Wit,hep-ph/0006023, ``Charge fluctuations in a final state with QGP,''hep-ph/0101258, A.~Bialas, arXiv:hep-ph/0203047.}{\small \par}
\bibitem{Capella:1989gc}{\small A.~Capella, C.~Merino and J.~Tran Thanh Van, 
Z.\ Phys.\ C {\bf 43}, 663 (1989).}{\small \par}
\bibitem{phenix02}{\small K.~Adcox [PHENIX Collaboration], arXiv:nucl-ex/0203014. }{\small \par}
\bibitem {voloshin}{\small S.~Mrowczynski, Phys.\  Rev.\  C { \bf 66}, 024904 (2002), 
C.~Pruneau, S.~Gavin and S.~Voloshin, Phys. \  Rev.\  C {\bf 66}, 044904 (2002)
[arXiv:nucl-ex/0204011]}.

\end{thebibliography}
\end{document}